\documentstyle[aps]{revtex}

\begin{document}
\input epsf

\title{TREATMENT OF THE QCD COUPLING IN HIGH ENERGY PROCESSES}

\author{B.I.~Ermolaev \\
 A.F.~Ioffe Physico-Technical Institute,
 St.Petersburg 194021, Russia\\
M.~Greco\\
Dipartimento di Fisica and INFN, University of Rome III, Italy\\
and\\
S.I.~Troyan\\
St.Petersburg Institute of Nuclear Physics, Gatchina,
St.Petersburg 188300, Russia}

\maketitle

\begin{abstract}
The treatment of the running QCD coupling in evolution equations is discussed.
It is shown that the use of the virtuality of ladder (vertical)
partons
 as the scale
for QCD coupling in every rung of ladder graphs is an approximation
that holds for DIS at large $x$ only. On the contrary, in the
small $x$ region the
coupling depends on the virtuality of $s$ -channel (horizontal) gluons.
This observation leads to different results for the Regge-like
processes and DIS structure functions at small $x$.
\end{abstract}

\section{INTRODUCTION}

In perturbative QCD, calculations of physical
quantities such as  scattering
amplitudes or  cross sections beyond the Born approximation
basically  requires one to account  for running $\alpha_s$ effects.
In fixed-order calculations, $\alpha_s$ becomes running
in a
straightforward way, by direct calculation of Feynman graphs.
However, in various approaches,
scattering amplitudes or cross sections are often calculated to all orders
of the perturbation theory. Generally this
means that some equations are constructed and solved and
when such equations include the  running  of $\alpha_s$,  the argument
of  $\alpha_s$  is usually fixed by
certain prescriptions which are based on different approximations.
For example, for hard QCD processes exploiting DGLAP evolution \cite{dglap},
one takes

\begin{equation}
\label{k}
\alpha_s = \alpha_s (k^2_{\perp})
\end{equation}
in every rung of quark/gluon ladder,
with $k_{\perp}$ being the transverse momentum of the ladder parton.
As a consequence, in the evolution equations

\begin{equation}
\alpha_s = \alpha_s(Q^2),
\label{q}
\end{equation}
with $\sqrt{Q^2}$ being the
upper limit for integration
over transverse momenta $k_{\perp}$ of ladder partons.
In particular, when the DGLAP is used for calculation of the structure
functions of the Deep Inelastic Scattering (DIS), $-Q^2$ is the virtuality
of the photon. The dependence of $\alpha_s$
given by Eqs~(\ref{k},\ref{q}) follows from the results of ref. \cite{abcmv}
for the treatment of $\alpha_s$ in hard QCD processes. 
On the other hand, some important
high energy QCD processes are  Regge-like, rather than hard. In particular,
 high energy forward and backward scattering and
DIS at small $x$ are typical Regge-like processes. The use of the hard -like prescription \cite{abcmv}
of the QCD coupling
 for these Regge processes should have been
justified somehow. Indeed,
much recent work uses the argument of $\alpha_s$
given  by  Eqs.~(\ref{k}, \ref{q}) without any comments.
For example, in ref \cite{thorne}
Eq.~(\ref{k}) is used
for incorporating running $\alpha_s$ into the BFKL equation
though it is well-known that the BFKL is constructed especially
for the Regge kinematics. Also,
works
\cite{c} suggest  $\alpha_s = \alpha_s(Q^2)$ for the polarised
DIS structure function
$g_1$ at small $x$. On the other hand, instead of exploiting
directly Eq.~(\ref{k}) refs \cite{emr}, \cite{m} prescribe a 
modified parametrisation of $\alpha_s$, namely

\begin{equation}
\alpha_s = \alpha_s(k^2_{\perp}/\beta),
\label{sman}
\end{equation}
where $\beta$ is the parameter of the standard Sudakov representation of
the ladder momentum $k$
\footnote{momenta $p$ and $q$ must obey $2pq \gg p^2 \sim q^2 \sim 0$,
otherwise they are arbitrary}:

\begin{equation}
\label{sud}
k_{\mu} = \alpha q_{\mu} + \beta p_{\mu} + k_{\perp \mu}~.
\end{equation}

Contrary  to refs. \cite{dglap} -\cite{m}, we have suggested
 \cite{egt} a quite different treatment of  $\alpha_s$ in the
evolution equations for the non-singlet structure functions at small $x$.
It is interesting that when the argument of $\alpha_s$ is discussed in
ref \cite{ciaf}, the first assumptions about the argument look similar to
Eqs.~(\ref{k}) though final result agrees
rather with \cite{egt} than with
Eq.~(\ref{k}).
In the present work we extend the treatment of  $\alpha_s$ suggested
 in \cite{egt} for the non-singlet structure functions, which are based on
 a quark ladder, to the the more complicated case of gluon ladders,
forming the base for the singlet structure functions of DIS.
We show that the treatment of $\alpha_s$ suggested in
refs. \cite{abcmv} for hard processes and often used subsequently
for description of different high energy QCD processes is an approximation
valid only for hard QCD processes. This treatment cannot be used
for the Regge kinematics.
We show  that the  $\alpha_s$ -dependence
given by Eq.~(\ref{k}) is valid only for hard
QCD processes where values of all
Mandelstam variables are of the same order.
In particular, for DIS it means
that  $\alpha_s = \alpha_s(Q^2)$ (or  $\alpha_s(k^2)$)
in evolution equations only when
$x \sim 1$.
On the contrary, the treatment of $\alpha_s$ of Ref.~ \cite{egt} is valid
for the  evolution of DIS structure functions at small $x$ and  for other
high energy QCD processes in the Regge
kinematics.
 The paper is organised as follows: in sect~2 we discuss the treatment of
  $\alpha_s$ in elementary parton processes. In sect.~3 we
extend it to partonic ladders in the Regge kinematics
and compare our results with results those of ref.~\cite{abcmv}. 
Finally, sect.~4 contains our concluding remarks.

\section{Incorporating running $\alpha_s$ into the Born
scattering amplitudes}

In this section we discuss the  argument of $\alpha_s$ in elementary
parton processes. In order to avoid unnecessary complications we regard
all these partons as slightly off-shell. We specify this fact explicitly in
Eq.~(\ref{ima}).
Let us consider first the scattering amplitude $M_{qq}$ for the process
of annihilation of a quark- antiquark pair into a quark -antiquark pair
of the same flavour:

\begin{equation}
\label{quarks}
q(p_1) + \bar{q}(p_2) \rightarrow q(p_1') + \bar{q}(p_2')
\end{equation}
in the forward kinematical region

\begin{equation}
\label{region}
s = (p_1 + p_2)^2 \approx -u = (p_2' - p_1)^2 \gg -t = -(p_1' - p_1)^2~.
\end{equation}

In the Born approximation $M_{qq}^{Born}$ is given by contributions of three
Feynman graphs. In each such graph the  intermediate gluon has
  momentum $k$, so that
 either $k^2 = s$ or $k^2 = u$ or at last $k^2 = t$. Such intermediate
gluons are  often addressed as $s$, $u$ or $t$ -channel gluons respectively.

According to the Optical theorem
\begin{equation}
\label{opt}
\sigma_{tot} \sim s^{-1} \Im_s M_{|_{t = 0}}~,
\end{equation}
so only the graph
with an intermediate $s$ -channel gluon in $M_{qq}^{Born}$
contributes to the total cross
section of the process. As our final goal is cross sections,  we consider only
this graph and denote its contribution  by $\tilde{M}_{qq}^{Born}$.
Adding the part of radiative corrections to $\tilde{M}_{qq}^{Born}$ which
makes $\alpha_s$ running $\tilde{M}_{qq}^{Born}$ transforms into

\begin{equation}
\label{mq}
\tilde{M}_{qq} = -4\pi\alpha_s(s)\frac{\bar{u}_(p_2)\gamma_{\mu}u(p_1)
\bar{u}_(p_2)\gamma_{\mu}u(p_1)}{s + \imath\epsilon} ~ .
\end{equation}
where $\alpha_s$ depends on $s$. In Eq.~(\ref{mq}),  we have dropped
the factor $4\pi$ and the colour structure.
The Feynman graphs corresponding to $\tilde{M}_{qq}$ are depicted in 
Fig.~1.  
The well-known BLM  \cite{blm} arguments allow one to consider only 
quark loop contributions to $\alpha_s$ and it is easy to show that 
the argument of 
$\alpha_s$  in Eq.~(\ref{mq}) is indeed $s$. To this end let us note first 
that at high energies and for 
space-like $q$ obeying

\begin{equation}
\label{lambda}
\Lambda^2 \equiv \Lambda^2_{QCD} <<-q^2
\end{equation}

$\alpha_s$ is given by

\begin{equation}
\label{alpha}
\alpha_s(q^2) = \frac{1}{b\ln(-q^2/\Lambda^2)} ~, \qquad
b =(11N - 2 n_f)/(12\pi)~.
\end{equation}

 As is well-known,
in Eq.~(\ref{alpha}) the contribution proportional to the number of
colours, $N$ comes from graphs with gluon loops whereas
the contribution proportional to $n_f$
comes from the
QED -like graphs where the gluon propagator is saturated only by
quark loops. Obviously, the main, logarithmic contribution of
quark loops is
proportional to $\ln(s)$.
As eventually quark and gluon loop contributions combine into
 Eq.~(\ref{alpha}), the argument of $\alpha_s$ in  Eq.~(\ref{mq}) is
 indeed $s$.

Now let us repeat the above reasoning, replacing the quarks by gluons,
i.e. let us consider the forward scattering process

\begin{equation}
\label{gluons}
g(p_1) + g(p_2) \rightarrow g(p_1') + g(p_2')
\end{equation}
in the forward region  Eq.~(\ref{region}).
We denote  the Born amplitude for the forward scattering
of gluons by $M_{gg}^{Born}$ and, similarly to Eq.~(\ref{mq}), denote
 the Born approximation for
the part of the forward  amplitude  with  $s$ -channel
intermediate gluon   by $\tilde{M}_{gg}^{Born}$ (see Fig.~2).
As is well-known, the Slavnov-Taylor~\cite{st}
identities state that accounting
for radiative corrections to $\tilde{M}_{gg}^{Born}$ lead to the same
dependency, $\alpha_s =\alpha_s(s)$. Therefore both for the quark and
gluon scattering,  adding radiative corrections to the part of the
Born scattering  amplitude with an $s$ -channel gluon
transforms it into

\begin{equation}
\label{mr}
\tilde{M} = -4\pi \alpha_s(s)\frac{R(s)}{s + \imath \epsilon}
\end{equation}
where we have dropped the color factors. $R$ includes vertices and spin
dependence. For example, for the scattering of on-shell
quarks,
$R(s) = R_q(s) = 2s$. For the gluon scattering, $R = R_g$
depends on polarisations  of gluons.
We consider only its part, $\tilde{M}$, with $s$ -channel gluon,
instead of the whole scattering amplitude $M$, disregarding parts of
$M$ with the $t$ and $u$ -channel gluons because  only $\tilde{M}$ has
non-zero imaginary part with respect to $s$.
Eq.~(\ref{mr}) than leads to

\begin{equation}
\label{im}
\Im_s M \equiv \Im\tilde{M} =
-\Im \alpha_s(s)\left(\frac{4\pi R(s)}{s}\right) +
\Re\alpha_s(s) \pi\delta(s)(4\pi R(s))
\end{equation}
where we have used $\Im R(s)= 0$. 
We remind the reader that throughout this paper
we assume all partons to be slightly off-shell. For the on-shell quarks,
the second term in the rhs of Eq.~(\ref{im}) is zero. It is easy to see that
$\alpha_s(s)$,  given by
Eq.~(\ref{alpha}) when its argument  is space-like,
can be rewritten as

\begin{equation}
\label{ima}
\alpha_s(s) = \frac{1}{b\ln(-s/\Lambda^2)} =
\frac{\ln(s/\Lambda^2) + \imath \pi}{b[\ln^2(s/\Lambda^2) + \pi^2]}~,
\end{equation}
 when  $s$ is time-like.
Indeed, any scattering amplitude $M(s)$ has a non-zero imaginary
part in $s$ when $s$ is positive. In particular when
$M$ depends on $s$ through logarithms, it means that

\begin{equation}
\label{ln}
M = M(\ln(-s))
\end{equation}
when $s$ is positive. However, Eq.~(\ref{ln}) does not fix the phase of
$-s=s e^{\imath \phi}$.
In order to fix $\phi$, one can use the
fact that $M$ does not have an imaginary part when $s$ is negative.
Then, the analytical continuation  of $M$ from positive
$s$ to negative $s$ is defined through
the upper path, because the cut in the $s$ -plane when
$s> 0$. In doing so, $s$ acquires
the factor $e^{\imath \pi}$ leading to
$\phi = -\pi$. Thus, instead of  Eq.~(\ref{ln})
one has

\begin{equation}
\label{lns}
M = M(\ln(s)-\imath \pi)
\end{equation}
when $s$ is positive.
Combining now Eq.~(\ref{im}) and Eq.~(\ref{ima}), we obtain

\begin{equation}
\label{imm}
\Im\tilde{M} = \frac{4\pi^2 R(s)}{b[\ln^2(s/\Lambda^2) + \pi^2]}
\left[\ln(s/\Lambda^2)\delta(s) - \frac{1}{s}  \right]~.
\end{equation}

Taking $\Im_sM$ corresponds to cutting all intermediate states in the  $s$
-channel and summing up such contributions.
The first term in the squared brackets in Eq.~(\ref{imm}) corresponds to the
case when the $s$ -channel intermediate state is an on-shell gluon whereas
the second term corresponds to many on-shell partons in the intermediate
state. Using the
$\delta$-function in Eq.~(\ref{imm}) when integrating the
first term over $s$  leads to a result which is
singular in the infrared region. The reason for that is
obvious: one can apply Eq.~(\ref{alpha}) for $\alpha_s(s)$ only when
$s\gg \Lambda^2$. In order to fix this we replace $\delta(s)$ by
$\delta(s - \mu^2)$ in Eq.~(\ref{mr}), with the infrared cut-off
$\mu \gg \Lambda$, arriving
therefore at the expression

\begin{equation}
\label{imu}
\Im\tilde{M} = \frac{4\pi^2 R(s)}{b[\ln^2(s/\Lambda^2) + \pi^2]}
\left[\ln(s/\Lambda^2)\delta(s - \mu^2) -
\frac{1}{s}  \right]
\end{equation}
for $s$ -channel imaginary part of $M_{qq}$ and $M_{gg}$ .

Finally let us consider the quark -gluon scattering, with both gluons
off-shell, $k^2 = (k')^2 <0$ :

\begin{equation}
\label{quarksgluons}
q(p) + g(k) \rightarrow q(p') + g(k')
\end{equation}
in the forward region (\ref{region}).
 In the Born approximation the only
Feynman graph with non-zero imaginary part in $s$ is depicted in Fig.~3.
Incorporating the radiative corrections to this graph in the same way as
 was done for quark-quark
and gluon-gluon scattering and using the BLM approach \cite{blm},
we conclude that for $qg$- scattering

\begin{equation}
\label{alphaqg}
\alpha_s = \alpha_s(k^2)~,
\end{equation}
which agrees with the treatment of $\alpha_s$ suggested in
ref. \cite{abcmv}.  Obviously, $\Im_s \alpha_s(k^2) = 0$ and therefore

\begin{equation}
\label{imuqg}
\Im\tilde{M_{qg}} =  \frac{4\pi^2 R(s)}{b~\ln(s/\Lambda^2)}
\delta(s - \mu^2)
\end{equation}
instead of Eq.~(\ref{imu}).

\section{QCD couplings in ladder graphs}

Now let us allow all partons in Eqs.~(\ref{quarks}),(\ref{gluons}) and
(\ref{quarksgluons}) to be off-shell so that their virtualities cannot
be neglected.
It is obvious that letting the on-shell partons which were  considered
in sect.~2  be off-shell 
does not change our conclusions concerning the argument of $\alpha_s$
in these scattering processes.
Our formulae Eqs.~(\ref{imu}),(\ref{imuqg}) also remain valid for off-shell
parton scattering, save a change of $R$ .
However, letting the partons be off-shell converts the amplitudes
we have considered into the rungs of partonic ladder graphs. Therefore
we conclude that incorporating the running QCD
coupling into the gluon-gluon and quark-quark rungs
fixes  the time-like virtuality of the intermediate $s$ -channel gluon
(the horizontal gluon line in Figs.~1-4) as the scale of $\alpha_s$.
In contrast to this, the scale of  $\alpha_s$ for the
quark-gluon and gluon-quark rungs is the space-like virtuality of the vertical
gluon lines.\\
Therefore the integrand of the quark ladder depicted in Fig.~4
contains $n + 1$  QCD couplings in such a way:

\begin{equation}
\label{quarklad}
\alpha_s((p_1 - k_1)^2)\alpha_s((k_1 - k_2)^2)
\alpha_s((k_2 - k_3)^2)....\alpha_s((p_2 + k_n)^2)~.
\end{equation}

When quark lines in Fig.~4 are replaced by the gluon lines,
the integrand of such purely gluon ladder
(we keep the same notations for the gluon ladder
momenta as for the quark ones) contains the same succession of
QCD couplings with the same
arguments as Eq.~(\ref{quarklad}). However, some interesting
objects
(e.g. for the singlet DIS structure function $g_1$) are made of mixing of
the quark and gluon rungs as shown in Fig.~5, so that
the integrand corresponding to Fig.~5 is proportional to
\begin{equation}
\label{mix}
\alpha_s(k_1^2)\alpha_s((k_1')^2)
\alpha_s((k_3 - k_4)^2)\dots\alpha_s(k_n^2)\alpha_s^2((k_n')^2)
\end{equation}
instead of the DGLAP succession\footnote{for the ladder parton momenta
$k^2_i \approx -k^2_{i~\perp}$}

\begin{equation}
\label{dglapmix}
\alpha_s(k_1^2)\alpha_s(k_2^2)\alpha_s(k_3^2)....\alpha_s^2(k_n^2) ~.
\end{equation}

A further difference concerning treatment of $\alpha_s$ arises when those
ladders are used for calculation of cross sections (e.g. DIS
structure functions), which contain the imaginary
part, $\Im_s A$ of these ladders: as all $s$ -channel gluons
(the horizontal lines in Fig.~5) are time-like,
the related $\alpha_s$ have non-zero  $\Im_s \alpha_s$ too.
Accounting for them makes the pattern  much more complicated than the
standard DGLAP approach.

Ladder graphs incorporating all orders in $\alpha_s$ are essential ingredient
of the DIS structure functions. We demonstrate now how and when
it is possible to justify
 the DGLAP-like argument (\ref{k}) of   $\alpha_s$ for them.  
To this aim let us suppose that some DIS structure function $\Phi$ obeys a 
Bethe-Salpeter equation. The part of this equation involving the 
$s$ -channel  cut of the ladder virtual gluon is represented by the
graph in Fig.~6. For example,  it corresponds to 
the DGLAP equation or  (after replacing the photon lines in Fig. 6 
by gluon ones) - to  
the first term of the BFKL equation with running $\alpha_s$. 
After simplification of the spin structure, 
its contribution, $\Phi(s, Q^2)$ can be written as follows : 

\begin{equation}
\label{phi}
\Phi(s, Q^2) =  (p + q)^2 \int\frac{d^4 k}{(2\pi)^4}
\frac{\Phi((q + k)^2, Q^2, k^2)~4 k^2_{\perp}}
{(k^2)^2~(q + k)^2} 
\Im_s 4 \pi \frac{\alpha_s((p - k)^2)}{(p - k)^2}
\end{equation}
where we have used the standard notations: $q$ stands for the virtual
photon momentum, $Q^2 = -q^2 > 0$ and
$k_{\perp}$ is transverse to
the plane formed by $q$ and $p$. The intermediate partons with momenta
$k$ in Fig.~6 can be either quarks or gluons. The external 
parton with momentum $p$ is assumed to be on-shell and we will neglect 
its virtuality, 
which is not important for our conclusions. 
As we are discussing the treatment of $\alpha_s$, we have dropped
unessential numerical  factors from rhs of Eq.~(\ref{phi}) and 
have used the same notation $\Phi$ 
both for the on-shell structure function in the lhs of Eq.~(\ref{phi})
and  for the off-shell one in the rhs. $\Phi$ is supposed to be a 
logarithmic function of $s$.   
In contrast to the conventional DGLAP or  BFKL cases, the rhs
of Eq.~(\ref{phi}) contains  $\Im \alpha_s$ instead of
  $\alpha_s$. Also, the argument of  $\alpha_s$ is the
virtuality of the horizontal gluon.
 However under certain conditions, one can rewrite
Eq.~(\ref{phi}) in the form containing  
$\alpha_s$ with  the DGLAP argument $k^2_{\perp}$.   To show this let us 
use the Sudakov
representation for $k$

\begin{equation}
\label{sudak}
k_{\mu} = \alpha (q +x p)_{\mu} + \beta p_{\mu} + k_{\perp \mu}~.
\end{equation}
where $x = Q^2/s $, $s \equiv 2pq$, and then introduce  the virtuality
$m^2 = (p - k)^2$ of the $s$ -channel gluon as a new variable
instead of $\alpha$.  As $k^2 = -(\beta m^2 + k^2_{\perp})/(1 - \beta)$, 
~$(q + k)^2 = 
[(\beta - x)[s(1 - \beta) -m^2] - k^2_{\perp}(1 - x)]/(1 - \beta)$, 
we arrive at 

\begin{eqnarray}
\label{phim}
\Phi(s, Q^2) = 
\frac{1}{ \pi^2}
\int dk^2_{\perp} d \beta d m^2 \frac{k^2_{\perp}}
{[\beta m^2 + k^2_{\perp}]^2}                 
\Phi\Big(\frac{[(\beta - x)[s(1 - \beta) -m^2] - k^2_{\perp}(1 - x)]}
{(1 - \beta)},  \nonumber \\ ~Q^2, ~ 
\frac{(\beta m^2 + k^2_{\perp})}{(1 - \beta)} \Big)  
\frac{s(1 - \beta)^2}
{[(\beta - x)[s(1 - \beta) -m^2] - k^2_{\perp}(1 - x)]}
\Im_s \frac{\alpha_s(m^2)}{m^2}~.~~~~~~
\end{eqnarray}

The integration over $\beta$ is supposed to yield logarithmic 
contributions. It is necessary for BFKL and important also for DGLAP 
when applied in the small- $x$ region.   
To this accuracy the integration over $\beta$ in Eq.~(\ref{phim}) 
  can  be written as $d\beta/\beta$, the main  contribution coming from 
 the region  

\begin{equation}
\label{beta}
1 \gg \beta \geq  \beta_{min}~,  
\end{equation}
with 

\begin{equation}
\label{betamin}
\beta_{min} =  x  + \frac{k^2_{\perp}(1 - x)}{ s-m^2} 
\end{equation}
as  the lowest 
limit of the integration. Therefore Eq.~(\ref{phim}) becomes

\begin{eqnarray}
\label{phimbetamin}
\Phi \approx \frac{1}{\pi^2}
\int_0^{\infty} d^2k_{\perp}
\int_{\beta_{min}}^{1} \frac{d\beta}{\beta} 
\int_0^{\infty}d m^2 
\Phi \Big(\beta(s - m^2) - k^2_{\perp}(1 - x), 
Q^2, ~(\beta m^2 + k^2_{\perp})\Big)    \nonumber \\            
\frac{s}{(s - m^2)} \frac{k^2_{\perp}}{[\beta m^2 + k^2_{\perp}]^2} 
\Im \Big(\frac{\alpha_s(m^2)}{m^2} \Big)~.~~~~~~~~~~
\end{eqnarray}

When the dependence of $\beta_{min}$ and   $\Phi$ upon  $m^2$ in the rhs 
of 
Eq.~(\ref{phimbetamin})  
 can be neglected, one can 
interpret the integration 
over $m^2$ as a dispersion integral for $\alpha_s$: 

\begin{eqnarray}
\label{intm}
\frac{1}{\pi}
\int_0^{\infty} d m^2 \frac{k^2_{\perp}}{[\beta m^2 + k^2_{\perp}]^2}
\Im \Big(\frac{\alpha_s(m^2)}{m^2} \Big) \approx \nonumber \\
\frac{1}{\pi}\int_0^{\infty} d m^2 \frac{1}{[\beta m^2 + k^2_{\perp}]}
\Im \Big(\frac{\alpha_s(m^2)}{m^2} \Big) \approx
\frac{\alpha_s(- k^2_{\perp}/\beta)}{ k^2_{\perp}}~. 
\end{eqnarray}

leading to an explicit $\alpha_s$ dependence. In particular
this is possible when $\Phi$ in the integrand of Eq.~(\ref{phimbetamin}) 
does not depend on $k^2$ and when $x$ is big enough to satisfy  

\begin{equation}
\label{dglap} 
\beta_{min} \approx x \gg  k^2_{\perp}(1 - x)/s .    
\end{equation}
Therefore the argument  $m^2$ of $\alpha_s$ in this kinematical 
condition is 
\begin{equation}
\label{argdglap} 
-k^2_{\perp}/x \leq m^2 = -k^2_{\perp}/\beta \leq -k^2_{\perp}
\end{equation}

Then  the integration over $\beta$ is in the region 
$x \leq \beta \ll 1$ and Eq.~(\ref{dglap}) is equivalent to the condition 
$k^2_{\perp}(1 - x) < sx = Q^2$ which is  exactly the DGLAP region.  
  According to Eq.~(\ref{dglap}), as 
$x$ must not be small, the dependence $\alpha_s \approx 
\alpha_s(k^2_{\perp})$ of  
Eq.~(\ref{argdglap}) actually coincides with the 
parametrisation suggested in Ref \cite{abcmv}. 

On the other hand, when $x \to 0$, $\beta_{min}$ depends on $m^2$:

\begin{equation}
\label{smallx}
\beta_{min}\approx \frac{ k^2_{\perp}}{s - m^2}~
\end{equation}

and, as well known,  $\Phi$ in the rhs of  Eq.~(\ref{phimbetamin}) 
depends 
on $k^2$. 
This makes impossible the use of 
Eq.~(\ref{intm}) for performing the integration over $m^2$. However, 
to the leading logarithmic accuracy, when 
$m^2 \ll s$ and $|k^2| \approx k^2_{\perp} \gg \beta m^2$, one can 
still perform the
integration in Eq.~(\ref{intm}),  without using the 
 dispersion relation, as:  

\begin{eqnarray}
\label{intmnew}
\frac{1}{\pi}
\int_0^{\infty} d m^2 \frac{k^2_{\perp}}{[\beta m^2 + k^2_{\perp}]^2}
\Im \Big(\frac{\alpha_s(m^2)}{m^2} \Big) \approx 
\frac{1}{ k^2_{\perp}}R,  \nonumber \\
R \equiv 
\frac{1}{\pi}\int_{\Lambda^2}^{k^2_{\perp}/\beta} 
\frac{d m^2} {m^2}
\Im{\alpha_s(m^2)} = 
\frac{1}{b} \int_0^{\ln( k^2_{\perp}/\beta \Lambda^2)}
\frac{d z}{z^2 + \pi^2} 
\end{eqnarray}
where we have used Eq.~(\ref{ima}) for $\Im \alpha_s$. 
Let us note that, contrary to Eq.~(\ref{intm}), the argument of 
$\alpha_s$ in Eq.~(\ref{intmnew}) is time-like. 
Introducing the infrared cutoff $\mu^2$ we arrive at the result
 
\begin{eqnarray}
\label{phismallx}
\Phi \approx \frac{1}{\pi} 
\int_{\mu^2/s}^1\frac{d\beta}{\beta} 
\int_{\mu^2}^{s\beta} \frac{dk^2_{\perp}}{ k^2_{\perp}}
\Phi \Big(\beta,  
Q^2, ~k^2_{\perp}\Big) R( k^2_{\perp}/\beta) 
\end{eqnarray}
for  $x \to 0$, which actually reproduces the first term of the BFKL for 
the 
case when the gluon with momentum $p$ is (nearly) on-shell. 
When $\pi^2$  
 is dropped, the integration in Eq.~(\ref{intmnew}) yields  
$R = \alpha_s(k^2_{\perp}/\beta)$ 
\footnote{A similar parameterization was used in Refs.~\cite{emr},\cite{m} for 
the non-singlet structure function.}, otherwise  
$R = (1/\pi b) \arctan[\ln(k^2_{\perp}/\beta \Lambda^2)/\pi]$~.
 
Eq.~(\ref{phismallx}) shows that   
one cannot drop the dependence on $\beta$ in $R$. 
In other words, there is no 
factorization between the dependence on the logitudinal and 
transverse momenta  at small $x$. 
  
A  simple way to solve such equations  
(and Eq.~(\ref{phi}) in particular) is  known from the 
Regge theory (see e.g. \cite{col}). 
The point is that the small- $x$ region is actually 
the Regge kinematical region where one can use the Mellin transform   
to simplify  equations similar to Eq.~(\ref{phi}). 
However, first one should introduce the signature amplitudes 
$M^{(\pm)} = [M(s, Q^2) \pm M(-s, Q^2)]/2$ so that the 
structure functions are proportional to 
$(1/\pi) \Im_s M^{(\pm)}$. Then one can use the 
Sommerfeld-Watson transform in the asymptotical form:  

\begin{equation}
\label{mellin}
M^{(\pm)}(s/m^2) = \int_{-\imath \infty}^{\imath \infty} 
\frac{d \omega}{2 \pi \imath} \Big(\frac{s}{m^2}\Big)^{\omega} 
\xi^{(\pm)}(\omega) F^{(\pm)}(\omega)~,
\end{equation}
where  $\xi^{(\pm)} = (e^{-\imath \pi \omega} \pm 1)/2$ 
are the signature factors well-known from the Regge theory.
Although the transform (\ref{mellin}) looks similar to the 
Mellin transform, actually there is a certain difference between them.
Indeed, contrary to the usual Mellin transform, the transform 
inverse to (\ref{mellin}) involves $\Im_s M^{(\pm)}$:  

\begin{equation}
\label{invmellin}
F^{(\pm)}(\omega) = \frac{-2}{\sin(\pi \omega)} \int_0^{\infty} 
d (s/m^2)(s/m^2)^{-1 - \omega} \Im_s M^{(\pm)}(s/m^2)~.
\end{equation}
 
This method can be easily used to solve the Bethe-Salpeter 
equation (\ref{phi}), with the argument of $\Im\alpha_s$ being  
time-like. 
An application of this method has been given in \cite{egt} for calculating 
the nonsinglet structure functions at small $x$. 
The fact that one cannot use the 
parametrisation $\alpha_s = \alpha_s(k^2_{\perp})$ for 
the DIS structure functions at small $x$ suggests in particular 
that in the generalisation of  
the BFKL equation, when accounting for   
running $\alpha_s$ effects,  the first term of the new kernel should  
contain  $\alpha_s$  with the time-like argument $k^2_{\perp}/\beta$ 
(at least in the region where $k^2_{\perp}/\beta$ dominates over the 
virtualities of 
the external gluons) whereas in the second term (corresponding to the 
virtual contribution)   $\alpha_s$ has 
the space-like argument $k^2_{\perp}$.   We will consider the 
running $\alpha_s$ effects for the BFKL in a forthcoming publication.

\section{Conclusion}

We have shown that when $\alpha_s$ is running in a general QCD process,  
providing  $\alpha_s$ with the argument
$k^2_{\perp}$ is not a general rule but  an approximation.
This approximation is valid for 
hard processes but fails for the Regge-like ones. It can be used
in the DGLAP evolution equations when they are applied
in the kinematical
region of large $x$ but cannot be used at small $x$.
In this regime we have suggested, for
both the quark-quark and the gluon-gluon rungs involving  $\Im \alpha_s$,
that the argument should be the virtuality of the $s$ -channel 
(horizontal) gluon. Being incorporated into evolution equations of the 
Bethe-Salpeter 
type, this leads to an effective coupling with the argument
  $k^2_{\perp}/\beta$.  
When the virtuality of this gluon is time-like,  $\pi^2$-terms
 appear in expression for $\Im \alpha_s$
(see Eqs.~(\ref{imm}),(\ref{ima})), due to the analytic properties of
$\alpha_s$.
 These  $\pi^2$-terms might be
quite important because of their big numerical value. In particular, they 
have an impact on the values of 
intercepts
of the non-singlet DIS structure functions~\cite{egt} and  are likely
to be important for the singlet structure functions as well.
However, when the starting point $\mu_0$ of evolution is big enouph, 
$\pi^2$ -terms   
can be easily neglected. It can be used for estimating $\mu_0$. 
In particular, we obtained\cite{egt} $\mu_0 \approx 5$GeV for the non-singlet 
structure functions.

\section{Acknowledgements}

We are grateful to G.~Altarelli, M.~Ciafaloini and D.A.~Ross for
interesting discussions of the results of this work.
The work is supported in part by grants INTAS-97-30494,
RFBR 00-15-96610 and by EU QCDNET contract FMRX-CT98-0194.

\newpage \centerline{\Large Figures}
\begin{figure}
\begin{center}
\begin{picture}(420,120)
\put(0,10){
\epsfbox{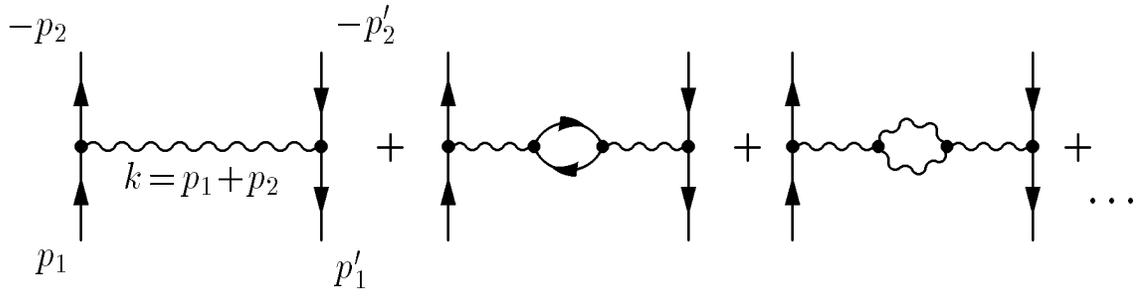}
}
\end{picture}
\end{center}
\caption{The Born amplitude $\tilde{M}_{qq}$, having imaginary part at
$s>0$, with running coupling included.}
\end{figure}
\begin{figure}
\begin{center}
\begin{picture}(120,120)
\put(0,10){
\epsfbox{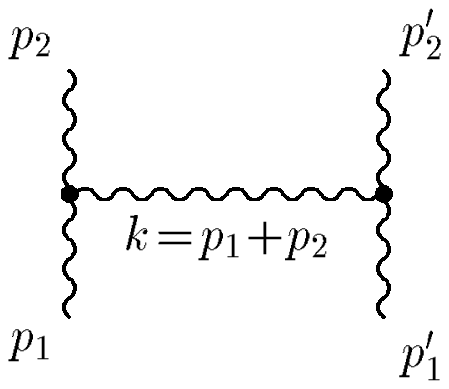}
}
\end{picture}
\end{center}
\caption{The Born amplitude $\tilde{M}_{gg}$, having imaginary part at
$s>0$.}
\end{figure}
\begin{figure}
\begin{center}
\begin{picture}(120,120)
\put(0,10){
\epsfbox{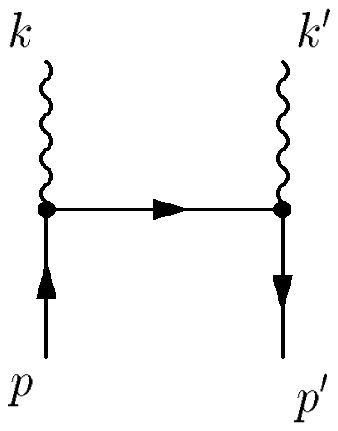}
}
\end{picture}
\end{center}
\caption{The Born amplitude $\tilde{M}_{qg}$, having imaginary part at
$s>0$.}
\end{figure}
\begin{figure}
\begin{center}
\begin{picture}(120,240)
\put(0,10){
\epsfbox{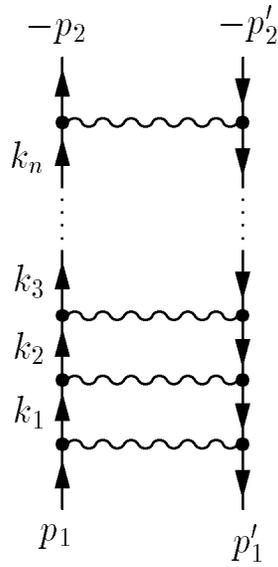}
}
\end{picture}
\end{center}
\caption{The quark ladder graph for $M_{qq}$.}
\end{figure}
\begin{figure}
\begin{center}
\begin{picture}(120,240)
\put(0,10){
\epsfbox{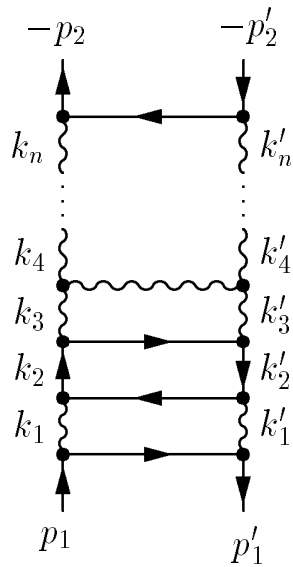}
}
\end{picture}
\end{center}
\caption{A general ladder graph for $M_{qq}$.}
\end{figure}
\begin{figure}
\begin{center}
\begin{picture}(120,240)
\put(0,10){
\epsfbox{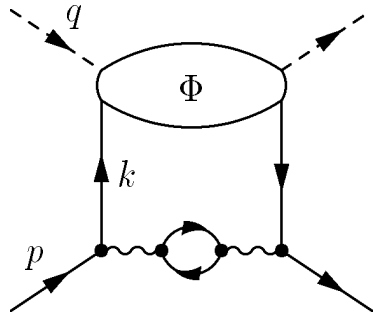}
}
\end{picture}
\end{center}
\caption{A graph for DGLAP evolution of a DIS structure function with
account of running couplig.}
\end{figure}

\end{document}